\documentclass[12pt]{article}

\usepackage{amsmath,amssymb,amsthm}
\usepackage[margin=2.5cm]{geometry}
\usepackage{hyperref}
\usepackage{color}
\usepackage{graphicx}
\usepackage{pgf}

\newcounter{Theorems}
\newcounter{Definitions}
\newcounter{Conjectures}
\begin{document}
\begin{titlepage}
\begin{flushright}

\end{flushright}

\begin{center}
{\Large\bf $ $ \\ $ $ \\
Deformations of AdS and exact sequences
}\\
\bigskip\bigskip\bigskip
{\large Andrei Mikhailov}
\\
\bigskip\bigskip
{\it Instituto de Fisica Teorica, Universidade Estadual Paulista\\
R. Dr. Bento Teobaldo Ferraz 271, 
Bloco II -- Barra Funda\\
CEP:01140-070 -- Sao Paulo, Brasil\\
}

\vskip 1cm
\end{center}

\begin{abstract}
We give examples of cohomologies of the superconformal algebra, relevant to computations in the AdS supergravity.
Our main examples are deformations of $AdS_5\times S^5$
transforming in finite-dimensional representations of the superconformal algebra
at the linearized level.
In the study of correlation functions, it is important to
compute the resolution of BRST-exact products of vertex operators.
The resolution is typically non-covariant, because of  cohomological obstacles.
Using the pure spinor formalism,
we develop a framework to describe these obstacles, and formulate conjectures about their structure.
\end{abstract}

\vfill
{\renewcommand{\arraystretch}{0.8}%
}

\end{titlepage}

\tableofcontents

\section{Introduction}\label{Introduction}

Global symmetries play fundamental role in AdS/CFT. In the case of $D=4$ $N=4$ supersymmetric Yang-Mills
theory, they form the  group of superconformal transformations $PSU(2,2|4)$.
The correlation functions are $PSU(2,2|4)$-invariant.
On the supergravity (SUGRA) side, this $PSU(2,2|4)$ is the group of super-isometries of $AdS_5\times S^5$.
Super-isometries are gauge transformations of SUGRA which preserve the undeformed  $AdS_5\times S^5$ background.
Infinitesimal deformations of $AdS_5\times S^5$ are linearized (around $AdS_5\times S^5$) solutions of SUGRA modulo
those which can be obtained as infinitesimal gauge transformations of $AdS_5\times S^5$.
The most important physical quantity is the  boundary S-matrix \cite{Witten:1998qj},
which is the exponential of the supergravity action with appropriate boundary conditions.
They are invariant under the gauge transformations modulo boundary terms arising from the Chern-Simons couplings.
The boundary conditions correspond to insertions of local operators on separate points on the boundary.
Therefore, the boundary terms arising from the Chern-Simons couplings vanish in the case of the boundary S-matrix,
leaving it $PSU(2,2|4)$-invariant. This fact is of crucial importance in the computations.
(The case of smeared, overlapping insertions appears to be more complicated, see \cite{Mikhailov:2019kfm} for preliminary considerations.)

If we have two states, transforming in two representations ${\cal H}_1$ and ${\cal H}_2$ of the symmetry algebra, we can ask which states
arise in the tensor product ${\cal H}_1\otimes {\cal H}_2$. This means that we look for invariant maps into another representation:
\begin{equation}\label{IntroInvariants}\mbox{Hom}_{\mathfrak{g}}({\cal H}_1\otimes {\cal H}_2, {\cal H}_3)\end{equation}
where $\mathfrak{g} = psu(2,2|4)$ and $\mbox{Hom}_{\mathfrak{g}}$ stands for $\mathfrak{g}$-invariant maps.
These invariant maps are very useful in the computation. When we compute the S-matrix, they control the poles in the scattering amplitudes.

A  generalization of symmetries are cohomology groups \cite{WeibelIntro}. For any representation $V$, we
can consider  the cohomology groups of $\mathfrak{g}$ ``with coefficients in $V$'', denoted $H^n(\mathfrak{g},V)$
(see Section \ref{sec:LieAlgebraCohomology} for a brief review of Lie algebra cohomology).

An important particular case is when $V$ is the space of linear maps between two representations $V_1$ and $V_2$.
In this case, $H^n(\mathfrak{g}, V)$ is called $\mbox{Ext}^n(V_1,V_2)$. Here ``Ext'' is an abbreviation for ``extension''.
In fact, $\mbox{Ext}^n(V_1,V_2)$ classify the exact sequences:
\begin{equation}0\longrightarrow V_2 \longrightarrow W_1\longrightarrow \ldots\longrightarrow W_n \longrightarrow V_1 \longrightarrow 0\end{equation}
modulo some equivalence relations.
The case $n=0$ corresponds to the $\mathfrak{g}$-invariant maps $\mbox{Hom}_{\mathfrak{g}}(V_1,V_2)$.
The case $n=1$ corresponds to the extensions of $V_1$ with $V_2$. In this case,
the action of $\xi\in\mathfrak{g}$ on $W_1$ can be written in the block-matrix form:
\begin{equation}\rho(\xi) = \left(\begin{array}{cc} \rho_{V_2}(\xi) & \psi(\xi) \\ 0 & \rho_{V_1}(\xi) \end{array}\right)\end{equation}
and the linear map $\psi$ defines a cocycle of $\mathfrak{g}$ with coefficients in $\mbox{Hom}(V_1,V_2)$.

In particular, we can consider, besides Eq. (\ref{IntroInvariants}), the higher invariants:
\begin{equation}\mbox{Ext}^n ({\cal H}_1\otimes {\cal H}_2, {\cal H}_3) = H^n(\mathfrak{g},\mbox{Hom}_{\mathbb{C}}({\cal H}_1\otimes {\cal H}_2, {\cal H}_3))\end{equation}
Given how useful are invariants, should we also consider the higher cohomologies of AdS representations?
In this paper we will give some examples of how the cohomologies, or Ext groups, are relevant in the context of pure
spinor superstring in $AdS_5\times S^5$.
Our examples relate to the theory of deformations of $AdS_5\times S^5$, which we will now briefly review.

\subsection{Vertex operators in the pure spinor formalism}\label{sec:VertexOperators}

In the pure spinor formalism the deformations of the background
correspond to the BRST cohomology in the ghost number two, and also in
the ghost number three \cite{Mikhailov:2014qka}. The ghost number three vertices can be obtained as products of
ghost number two vertices and ghost number one vertices (corresponding to the global symmetries).
Ghost number three cohomology is in the same representations as ghost number two.
(In the case of bosonic string, they are related by the rotational zero mode of the $b$-ghost: $V_2 = (b_0 - \bar{b}_0)V_3$.
    Integrating over rigid rotations of the insertion disk, we go from ghost number three to ghost number two.
    See \cite{Kishimoto:2024yuw} for a recent discussion of the ghost number three vertices for the bosonic string.)
The ghost number two vertex operators (but not
ghost number three) transform in the same representations of the
symmetry group as their cohomology.  This is a nontrivial fact proven
in \cite{Mikhailov:2011si}.  A priori this would be true only up to
BRST-exact terms. But it is possible to fix the freedom of choice of a
representative vertex in a cohomology class in such a way, that they
transform in the same representation as the cohomology. In this case
we say that the vertices ``transform covariantly''. This is special
to AdS, not true in flat space \cite{Mikhailov:2023fvx}.

\subsection{Extensions of representations}\label{sec:IntroExtensions}

It is often important to know the resolution of a BRST exact operator. For example, consider the product of
three ghost number two vertex operators:
\footnote{Here we consider classical supergravity, and classical worldsheet theory.
Eventually, it would be important to understand what happens at the quantum level, when we have to take
into account the regularization. We could start with taking the product $V_1(z_1,\bar{z}_1)V_2(z_2,\bar{z}_2)V_3(z_3,\bar{z}_3)$
of operators at three different points. For now, we just restrict ourselves to classical theory.
Moreover, since we only consider massless states, we can forget about the $(z,\bar{z})$-dependence altogether.
The massless vertex operators in the pure spinor formalism do not involve worldsheet derivatives.
They are some functions on a spin bundle over $AdS_5\times S^5$.}
\begin{equation}V_1V_2V_3\end{equation}
It is BRST exact, because the cohomology in the ghost number six is trivial.
(Here we do not impose any particular boundary conditions, which would limit the growth at infinity.
      Also, we do not impose any requirements of operators being primary fields.
      This is analogous in the case of bosonic string to allowing the vertex operators to depend on $X$ without
      derivatives \cite{Belopolsky:1995vi}, see Section \ref{sec:IntroBosonicString}.)
This means that exists some operator $W$ such that:
\begin{equation}\label{VVVisQW}V_1 V_2 V_3 = QW\end{equation}
Suppose that $V_1$, $V_2$ and $V_3$ transform in representations $H_1$, $H_2$ and $H_3$
of $PSU(2,2|4)$. Then $V_1V_2V_3$ transforms in the tensor product $H_1\otimes H_2\otimes H_3$.
The BRST operator $Q$ is $PSU(2,2|4)$-invariant. However, generally speaking, it is not true
that $W$ transforms in $H_1\otimes H_2\otimes H_3$. Rather, there is a nontrivial extension of $H_1\otimes H_2\otimes H_3$,
which we call $Y_5$. It fits into the exact sequence:
\begin{equation}0\longrightarrow X_5 \overset{\subset}{\longrightarrow} Y_5 \overset{Q}{\longrightarrow} H_1\otimes H_2\otimes H_3 \longrightarrow 0\end{equation}
where $X_5$ is some space of $Q$-closed operators of the ghost number five. Since $X_5$ has ghost number five, it is
in fact BRST-exact. Therefore we can continue:
\begin{align}
 & 0\longrightarrow X_4 \overset{\subset}{\longrightarrow} Y_4 \overset{Q}{\longrightarrow} X_5\longrightarrow 0 \\
 & 0\longrightarrow X_3 \overset{\subset}{\longrightarrow} Y_3 \overset{Q}{\longrightarrow} X_4\longrightarrow 0\end{align}
The space $X_3$ consists of $Q$-closed operators of the ghost number three.
The concatenation of these short exact sequences gives us the long exact sequence:
\begin{equation}\label{IntroSeqExt3}0\longrightarrow X_3\longrightarrow Y_3 \longrightarrow Y_4 \longrightarrow Y_5 \longrightarrow H_1\otimes H_2\otimes H_3\longrightarrow 0\end{equation}
and the corresponding cohomology class in $\mbox{Ext}^3\left(H_1\otimes H_2\otimes H_3,X_3\right)$.
There is a nontrivial cohomology
at the ghost number three. Projecting to the cohomology, we get an element
\begin{equation}\label{IntroExt3}\psi\in\mbox{Ext}^3\left(\;H_1\otimes H_2\otimes H_3 \;,\; H_Q^3 \;\right)\end{equation}
Such cohomology groups as in Eq. (\ref{IntroExt3}) should be important for  {\bf correlation functions}.
At this time, there is no satisfactory prescription for the  computation of the
correlation functions in AdS using the pure spinor formalism. We can guess, following \cite{Berkovits:2008ga},
that the correlation function can be computed as some kind of an integral of
the product of vertices (which are functions on the pure spinor bundle over AdS).
Consider the three-point correlation function:
\begin{equation}\langle V_1 V_2 V_3\rangle\end{equation}
Eq. (\ref{VVVisQW}) seems to imply that the correlation function is zero.
The reason for it to be nonzero would be that $W$ is not an admissible cochain.
If it transformed covariantly, it would have been admissible.
This is similar to the failure of the dilaton zero mode decoupling in the theory of bosonic string,
as we will now briefly review.

\subsection{Case of bosonic string}\label{sec:IntroBosonicString}

In the context of {\bf bosonic string in flat space}, it was discovered in \cite{Belopolsky:1995vi} that
the BRST exact state of the {\bf open bosonic string}:
\begin{equation}c\partial X^{\mu} = QX^{\mu}\end{equation}
does not decouple in the amplitude:
\begin{equation}\left\langle c\partial X^{\mu} ce^{ip_1X} ce^{ip_2X}\right\rangle \neq 0\end{equation}
The BRST-exact operator $c\partial X^{\mu}$ transforms in the vector representation $\mathbb{V}$ of the
Poincare group. In this representation, the translations $T_{\mu}$ act as zero, and the Lorentz rotations $L_{[\mu\nu]}$
rotate the index $\mu$ of $c\partial X^{\mu}$.
The resolution of the BRST-exact operator $c\partial X^{\mu}$ corresponds to the following extension:
\begin{equation}0\longrightarrow \mathbb{R} \longrightarrow \widehat{\mathbb{V}} \overset{\partial}{\longrightarrow} \mathbb{V} \longrightarrow 0\end{equation}
Here $\mathbb{V}$ consists of the expressions linear in $\partial X^{\mu}$ and $\widehat{\mathbb{V}}$ consists
of the expressions linear in $X^{\mu}$ (``bare $X$'') and constants.
The corresponding cocycle of the Poincaré algebra $\mathfrak{P}$ is:
\begin{align}
 & \Psi\;\in\;C^1(\mathfrak{P}, \mbox{Hom}(\mathbb{V},\mathbb{R}))\label{CocycleInFlatSpace} \\
 & \Psi\langle T_{\mu}\rangle \partial X^{\nu} \;=\;\delta^{\nu}_{\mu}\end{align}
The appearance of ``bare $X$'' is controlled by the cocycle Eq. (\ref{CocycleInFlatSpace}). Since it is a nontrivial
cocycle, it cannot be removed by choosing a different $Q^{-1}c\partial X$ (adding a $Q$-closed expression).
In bosonic string flat space background, this leads to the following peculiarity
in the definition of the correlation functions. The correlation functions (in the presence of ``bare $X$'') are
ill-defined {\it if we insist} that they are proportional to the delta-function of the momentum
conservation $\delta\left(\Sigma p_i\right)$. As was shown in \cite{Belopolsky:1995vi}, in the presence
of bare $X$, the {\it derivatives} of the
delta-function of the momenta appear, leading to the non-decoupling of the BRST-trivial expressions (if they contain bare $X$).

Similar considerations can be applied in general to a three-point function. Let us consider the correlation function of three
gluon vertices:
\begin{equation}\left\langle
\;(cE_{1\mu_1}\partial X^{\mu_1}e^{ip_1X})(z_1,\bar{z}_1)
\;(cE_{2\mu_2}\partial X^{\mu_2}e^{ip_2X})(z_2,\bar{z}_2)
\;(cE_{3\mu_3}\partial X^{\mu_3}e^{ip_3X})(z_3,\bar{z}_3)
\;
\right\rangle\end{equation}
Consider the limit $z_1\to z_2$. There are terms proportional to $c\partial c\over (z_1 - z_2)$, and to $c\partial^2 c$,
coming from the contraction $\left\langle\partial X(z_1) \partial X(z_2)\right\rangle$. They are BRST exact.
There are also terms proportional to $c\partial c$ coming from the contraction of $\partial X$ with $e^{ipX}$.
They are also BRST exact, {\it i.e.} of the form $Q{\cal O}_{12}$, but ${\cal O}_{12}$ includes the denominator
$1\over (p_1+p_2)^2$ singular when $(p_1 + p_2)^2=0$. Let $c{\cal R}_{12}e^{i(p_1+p_2)X}$ be the residue:
\begin{equation}{\cal O}_{12} = {c{\cal R}_{12}e^{i(p_1+p_2)X}\over (p_1 + p_2)^2} + \ldots\end{equation}
(With these notations ${\cal R}_{12}$ is an expression linear in $\partial X$.)
The correlation function becomes:
\begin{equation}\label{AfterMovingQ}\left\langle {c{\cal R}_{12}\over (p_1+p_2)^2} e^{i(p_1+p_2)X}
Q\left((cE_{3\mu_3}\partial X^{\mu_3}e^{ip_3X})(z_3,\bar{z}_3)\right)\right\rangle\end{equation}
The integration over the zero modes of $X$ leads to $\delta(p_1+p_2+p_3)$, while
$Q\left((cE_{3\mu_3}\partial X^{\mu_3}e^{ip_3X})(z_3,\bar{z}_3)\right) \simeq p_3^2$.
This $p_3^2$ cancels $1\over (p_1 + p_2)^2$ leading to a finite result.
In fact, ${\cal R}_{12}$ is  related to an Ext class.
To understand this, notice that the computation of $Q^{-1}(c\partial c \ldots)$ involves the inversion of the Maxwell kinetic operator,
essentially the inversion of the Laplace operator $\square$. We have to solve the equation:
\begin{equation}\label{InverseSquare}\square F = e^{ipX}\end{equation}
where $p = p_1 + p_2$. The solution is:
\begin{equation}\label{SingularSolution}F = - {e^{ipX}\over p^2}\end{equation}
with the singularity at $p^2=0$. This singularity is the obstacle to finding a covariant solution
of Eq. (\ref{InverseSquare})  for light-like $p$, and simultaneously the reason why the expression
in Eq. (\ref{AfterMovingQ}) is nonzero.
But there is a non-covariant solution, which exists precisely when $p^2=0$:
\begin{equation}\label{NonsingularSolution}\tilde{F} = {1\over 2i} {X^0\over p^0} e^{ipX}\end{equation}
This solution transforms in a nontrivial extension of the one-dimensional representation of the translation algebra,
generated by $e^{ipX}$. (This extension  consists of the
functions of the form $a(X)e^{ipX}$ where $a$ is a constant plus linear function of $X$.)
The relation between Eqs. (\ref{SingularSolution}) and (\ref{NonsingularSolution}) can be interpreted as follows.
Since $\square(\delta(p^2)e^{ipx})=0$, the following expression is a harmonic function:
\begin{equation}{1\over 2p^0}{\partial\over\partial p^0}
\left(
      \delta(p^2)e^{ipX}
      \right)
=
\delta(p^2)
\left(
      - {1\over p^2} + {i\over 2} {X^0\over p^0}
      \right)
e^{ipX}\end{equation}
This means that $f = -\delta(p^2){1\over p^2}e^{ipX}$ and $\tilde{f} =\delta(p^2){1\over 2i}{X^0\over p^0}e^{ipX}$ both solve the same equation:
\begin{equation}\square f = \square \tilde{f} = \delta(p^2) e^{ipX}\end{equation}
In other words, having a family of irreducible representations, we obtain non-semisimple representations by the differentiation
with respect to the parameter of the family.

\subsection{Superstring in $AdS_5\times S^5$}\label{sec:IntroAdS}

In AdS/CFT we study the boundary S-matrix, which corresponds to the correlations function
 in the QFT living on the boundary \cite{Witten:1998qj}.
For example, the three point function $\left\langle {\cal O}_1 {\cal O}_2 {\cal O}_3\right\rangle_{QFT}$.
There is no  momentum conservation delta-function. However, when covariance is broken
by a cohomological cocycle (nontrivial extension), we would guess that the integrand would be growing
in the noncompact directions of AdS, making the integral divergent. This, again, would lead to the non-decoupling
of the BRST-trivial state. The growth of a function in the non-compact directions of AdS is
determined by its symmetry transformations. If $W$ of Eq. (\ref{VVVisQW}) transformed in the same
representation $H_1\otimes H_2\otimes H_3$, then it would have grown just like $V_1V_2V_3$, and therefore
would be an admissible cochain, and the correlator would be zero.
At this time we do not have a precise statement about how to compute the correlation function from the known cocycle.
Moreover, the cocycle is not known.
The actual computations of the resolution of $Q$-exact expression and the corresponding cohomology class are very difficult
(although, unlike correlation functions, completely straightforward;
           see Section \ref{OpenQuestions}).
We have only been able to do them in one particular case, which we
consider in Section \ref{ToyExample}, where the actual resolution of a BRST-exact expression is known explicitly.
We will first describe the general idea, that a resolution generates a Lie superalgebra cohomology class. We then explicitly describe
this class in this particular case.
In the rest of this paper we will attempt to ``extrapolate'' the result of
Section \ref{ToyExample}
to the cases where the actual resolution is
not explicitly known.  This is mostly guesswork.
We will try to  keep track of the conjectures and always say when we are not sure.

\section{Brief review and notations}\label{Notations}

\subsection{Pure spinors in AdS}\label{sec:PureSpinorsInAdS}

We will define super-$AdS_5\times S^5$ as the coset space $H\backslash G$ where $G=PSU(2,2|4)$ and $H=SO(1,4)\times SO(5)$.
Let $\mathfrak{g}$ and $\mathfrak{h}$ denote the corresponding Lie superalgebras:
\begin{align}
\mathfrak{g}\;=\; & \mbox{Lie}\,G = psu(2,2|4) \\
\mathfrak{h}\;=\; & \mbox{Lie}\,H = so(1,4)\oplus so(5)\end{align}
Let $S$ denote the chiral spinor representation of $SO(1,9)$ restricted to $H=SO(1,4)\times SO(5)$. Since we are
working with type IIB theory, we need two copies of $S$ which we denote $S_L$ and $S_R$. We consider
the cone inside $S_L\oplus S_R$ which is parameterized by $\lambda_L\in S_L$ and $\lambda_R\in S_R$ with the following
constraints:
\begin{equation}\label{PureSpinorConstraint}\lambda_L^{\alpha} \Gamma^m_{\alpha\beta}\lambda_L^{\beta}
=
\lambda_R^{\hat{\alpha}}\Gamma^m_{\hat{\alpha}\hat{\beta}}\lambda_R^{\hat{\beta}}
= 0\end{equation}
Functions on $PSU(2,2|4)$ can be thought of as functions on $G$ satisfying:
\begin{equation}f(hg) = f(g)\end{equation}
for $g\in G$, $h\in H$. The pure spinor BRST complex of Type IIB superstring in $AdS_5\times S^5$ requires sections
of the pure spinor bundle, rather than functions. It is defined in the following way.
The space of cochains of the ghost number $n$ is the space of functions of $g$, $\lambda_L$, $\lambda_R$, of the total degree $n$
in $\lambda_L$ and $\lambda_R$, satisfying:
\begin{equation}\label{ConeSections}f(\;\mbox{Ad}_h\lambda_L\,,\;\mbox{Ad}_h\lambda_R\,,\;hg\;) = f(\lambda_L,\lambda_R,g)\end{equation}
The BRST operator acts as follows:
\begin{align}
Q \;=\; & \lambda^{\alpha}_L\nabla^L_{\alpha}  + \lambda^{\hat{\alpha}}_R\nabla^R_{\hat{\alpha}}\label{BRST} \\
\mbox{where \hspace{1ex}} & \nabla^L_{\alpha}f = {d\over d\epsilon} f(e^{\epsilon t^L_{\alpha}}g) \\
 & \nabla^R_{\hat{\alpha}}f = {d\over d\epsilon} f(e^{\epsilon t^R_{\hat{\alpha}}}g)\end{align}
Here we introduced a Grassmann odd parameter $\epsilon$, to define left regular action  of odd
generators $t_{\alpha}^L$ and $t_{\hat{\alpha}}^R$.
The vector field $Q$ is nilpotent.
Verification of nilpotence requires Eqs. (\ref{PureSpinorConstraint}) and (\ref{ConeSections}).

Let $P^n$ be the space of polynomials of the total degree $n$ of $\lambda_L$ and $\lambda_R$, a representation of $H$.
The appropriate notation for the space of cochains of the ghost number $n$ is:
\begin{equation}C^n = \mbox{Coind}_H^G P^n\end{equation}
--- the coinduced representation.

\subsection{Lie algebra cohomology}\label{sec:LieAlgebraCohomology}

The space of $n$-cochains of the Lie algebra cohomology complex \cite{GelfandManin}, \cite{Knapp}, \cite{WeibelIntro}
with the coefficients in a representation $V$ is defined as follows:
\begin{equation}C^n_{\rm Lie}\left(\mathfrak{g},V\right) = \mbox{Hom}_{\mathbb{C}}\left(\Lambda^n\mathfrak{g}, V\right)\end{equation}
In the case of Lie superalgebras, it is better to think of it as the space of functions (or just polynomial functions)
on the superspace $\Pi \mathfrak{g}$ (the $\mathfrak{g}$ with flipped statistics).
The coordinates on $\Pi \mathfrak{g}$ are called ``Faddeev-Popov ghosts'' $C^a$. The differential is:
\begin{equation}d_{\rm Lie} = C^a\rho_V(t_a) - {1\over 2}C^a C^b f_{ab}^c {\partial\over\partial C^c}\end{equation}

We will write $\mathfrak{g}$ instead of $\mbox{ad}_{\mathfrak{g}}$ for the adjoint representation of $\mathfrak{g}$. For example:
\begin{equation}H^1\left(\mathfrak{g},\mathfrak{g}\right)
=
H^1\left(\mathfrak{g},\mbox{ad}_{\mathfrak{g}}\right)\end{equation}
--- the derivations of $\mathfrak{g}$.

\section{Explicit example}\label{ToyExample}

We will first give the general explanation of the example, and then the actual computation.

\subsection{General explanation}\label{sec:GeneralScheme}

One can obtain ghost number three vertices as products of ghost number two vertices with the ghost
number one vertices. The ghost number one vertices correspond to global symmetries. For every symmetry
$x\in \mathfrak{g}$, there is the corresponding ghost number one vertex:
\begin{equation}\Lambda\langle x\rangle = \mbox{STr}\left(x g^{-1}(\lambda_L - \lambda_R)g\right)\end{equation}
Given a ghost number two vertex $V$, consider the product $\Lambda\langle x\rangle V$. It is nontrivial
in the BRST cohomology iff the ghost number two vertex $x.V$ is BRST-nontrivial,
where $x.V$ denotes the action of the infinitesimal symmetry $x$ on the vertex $V$ \cite{Mikhailov:2014qka}.

As a particular case let us take
\begin{equation}\label{DilatonZeroMode}V = \mbox{STr}(\lambda_L\lambda_R)\end{equation}
This is the ghost number two vertex corresponding to changing the AdS radius \cite{Berkovits:2008ga}.
Since $V$ is invariant under symmetries, and therefore $x.V=0$,  $\Lambda\langle x\rangle V$ is exact.
Therefore exists some $W$ such that $\Lambda\langle x\rangle V = Q(W)$. We want to study the properties of the
correspondence $\Lambda\langle x\rangle\,V\mapsto W$. First of all, let us view $\Lambda\langle x\rangle V$ as a function of $x\in\mathfrak{g}$;
let us call it $F$:
\begin{align}
F\;:\; & \mathfrak{g}\rightarrow \mbox{Coind}_{H}^{G} P^3 \\
F\langle x\rangle\;=\; & \Lambda\langle x\rangle V\end{align}
(We use the angular brackets for the arguments, {\it i.e.} $F\langle x\rangle$ instead of  $F(x)$, to
    stress that the dependence on $x$ is linear.)
Since $\Lambda\langle x\rangle V$ is exact, we can choose for every $x$ some $F'\langle x\rangle$ such that:
\begin{equation}F\langle x\rangle = QF'\langle x\rangle\end{equation}
Let us consider the variation of $F'\langle y\rangle$ under some symmetry $x\in \mathfrak{g}$. In particular, we observe:
\begin{equation}Q\left(x . F'\langle y\rangle - F'\langle [x,y]\rangle\right) = 0\end{equation}
This follows from the $\mathfrak{g}$-invariance of $Q$ and $x.F\langle y\rangle - F\langle [x,y]\rangle = 0$.
(Since $x.F\langle y\rangle - F\langle [x,y]\rangle = 0$, we say that $F$ is ``covariant''. But $F'$ is not covariant,
       it is only covariant up to a BRST-closed expression.)
Let us consider the BRST cohomology class:
\begin{align}
G\;\in\; & C^1\left(\mathfrak{g},\mbox{Hom}_{\mathbb{C}}\left(\mbox{ad}_{\mathfrak g},H^2_Q\right)\right) \\
G\langle x\rangle \langle y\rangle \;=\; & x . F'\langle y\rangle - F'\langle [x,y]\rangle  \quad\mbox{mod}\quad \mbox{im}\;Q\end{align}
By construction, this is annihilated by $d_{\rm Lie}$. Taking its Lie cohomology class, we get an element:
\begin{equation}\label{ResolutionOfLambdaStr}[G] \in H^1\left(\mathfrak{g},\mbox{Hom}_{\mathbb{C}}\left(\mbox{ad}_{\mathfrak g},H^2_Q\right)\right)\end{equation}
To summarize:

\begin{itemize}
\item The resolution of the BRST exact vertex $\Lambda\langle x\rangle \mbox{STr}(\lambda_L\lambda_R)$ generates
an element of the cohomology group $H^1\left(\mathfrak{g},\mbox{Hom}_{\mathbb{C}}\left(\mbox{ad}_{\mathfrak g},H^2_Q\right)\right)$

\end{itemize}

\subsection{Explicit computation}\label{sec:ExplicitComputation}

We will now explicitly compute the cohomology class defined in Eq. (\ref{ResolutionOfLambdaStr}).

It was proven in \cite{Bedoya:2010qz} that:
\begin{equation}\label{OldIdentity}\mbox{STr}(\lambda_3\lambda_1) (g^{-1}(\epsilon\lambda_3 - \epsilon\lambda_1)g)_a f^a_{bc}
\;=\;
\epsilon Q
\left(
      \mbox{STr}(\lambda_3\lambda_1)
      \mbox{STr}
      \left(
            (gt_bg^{-1})_{\bar{1}} ({\bf 1} - 2{\bf P}_{31}) (gt_cg^{-1})_{\bar{3}}
            \right)
      \right)\end{equation}
Here ${\bf P}_{31}$ is some projection operator. Its precise form is not important for us now.
It is a rational function of the pure spinor variables, such that $\mbox{STr}(\lambda_L\lambda_R){\bf P}_{31}$ is a polynomial.
Therefore, the right hand side of Eq. (\ref{OldIdentity}) is of the form $Q(\ldots)$ where $\ldots$ is a quadratic polynomial
in the pure spinor variables. In our present notations, Eq. (\ref{OldIdentity}) reads as follows. For any two elements of the
algebra $x\in\mathfrak{g}$ and $y\in\mathfrak{g}$:
\begin{align}
F\langle [x,y]\rangle\;=\; & Q \widetilde{F}(x,y)\label{QFtilde} \\
\mbox{where\hspace{1ex}} & \widetilde{F}(x,y)
=
{1\over 2}
\mbox{STr}(\lambda_L\lambda_R)
\mbox{STr}\left((gxg^{-1})_{\bar{1}}({\bf 1} - 2{\bf P}_{31})(gyg^{-1})_{\bar{3}} - (x\leftrightarrow y)\right)\label{DefFtilde}\end{align}
Notice that the map $\widetilde{F}$ defined in Eq. (\ref{DefFtilde}) is bilinear in its arguments.
Therefore we can consider it as a map from $\mathfrak{g}\wedge\mathfrak{g}$ to the space of vertex operators:
\begin{equation}\widetilde{F}\;:\;\mbox{ad}_{\mathfrak{g}}\wedge\mbox{ad}_{\mathfrak{g}} \rightarrow \mbox{Coind}_H^GP^2\end{equation}
This implies that for all $z$, $F\langle z\rangle$ is BRST-exact. This is because
any $z\in \mathfrak{g}$ can be represented as a commutator of some two elements of $\mathfrak{g}$,
{\it i.e.} $z = [x,y]$.
However, it is not possible to choose $x$ and $y$ as functions of $z$ in a $\mathfrak{g}$-covariant manner.
In other words, the short exact sequence of $\mathfrak{g}$-modules:
\begin{equation}\label{ExampleShortExactSequence}0
\longrightarrow
(\mbox{ad}_{\mathfrak{g}}\wedge\mbox{ad}_{\mathfrak{g}})_0
\longrightarrow
\mbox{ad}_{\mathfrak{g}}\wedge\mbox{ad}_{\mathfrak{g}}
\overset{[\_,\_]}\longrightarrow
\mbox{ad}_{\mathfrak{g}}
\longrightarrow
0\end{equation}
does not split. The best we can do is to choose some map $z\mapsto \sum_i x_i\wedge y_i$, linear but not $\mathfrak{g}$-invariant,
so that $\sum_i[x_i,y_i] = z$. Let us call this map $\sigma$:
\begin{align}
 & \sigma\;:\;\mbox{ad}_{\mathfrak{g}} \longrightarrow \mbox{ad}_{\mathfrak{g}}\wedge \mbox{ad}_{\mathfrak{g}} \\
 & [\_,\_]\circ \sigma = \mbox{id}\end{align}
Consider the  deviation of $\sigma$ from being $\mathfrak{g}$-invariant:
\begin{equation}\tau\langle C\rangle = \rho_{\mathfrak{g}\wedge\mathfrak{g}}\langle C\rangle\circ \sigma - \sigma \circ \rho_{\mathfrak{g}}\langle C\rangle\end{equation}
(Here $C$ is the Faddev-Popov ghost, see Section \ref{sec:LieAlgebraCohomology}.)
Notice that $\tau$ takes values in $(\mathfrak{g}\wedge\mathfrak{g})_0$. The cohomology class
\begin{equation}[\tau]\in H^1\left(\mathfrak{g}, \mbox{Hom}(\mbox{ad}_{\mathfrak{g}},(\mbox{ad}_{\mathfrak{g}}\wedge\mbox{ad}_{\mathfrak{g}})_0)\right)\end{equation}
does not depend on the choice of $\sigma$. Let us compose $\tau$ with $\widetilde{F}$:
\begin{equation}[\widetilde{F}\circ\tau] \in H^1\left(\mathfrak{g}, \mbox{Hom}\left(\mbox{ad}_{\mathfrak{g}},\mbox{Coind}_H^GP^2\right)\right)\end{equation}
Eq. (\ref{QFtilde}) implies that the restriction of $\widetilde{F}$ on
$(\mbox{ad}_{\mathfrak{g}}\wedge\mbox{ad}_{\mathfrak{g}})_0$ is $Q$-closed. Taking its BRST cohomology class, we get:
\begin{equation}[\widetilde{F}\circ\tau\;\mbox{mod im}\;Q] \in H^1\left(\mathfrak{g}, \mbox{Hom}\left(\mbox{ad}_{\mathfrak{g}},H^2_Q\right)\right)\end{equation}
Given $B\in (\mbox{ad}_{\mathfrak{g}}\wedge\mbox{ad}_{\mathfrak{g}})_0$, the BRST cohomology class of $\widetilde{F}\langle B\rangle$ has been
identified in \cite{Flores:2019dwr} as the infinitesimal beta-deformation of $AdS_5\times S^5$ with the parameter $B$.
Infinitesimal beta-deformations transform in:
\begin{equation}{(\mbox{ad}_{\mathfrak{g}}\wedge\mbox{ad}_{\mathfrak{g}})_0\over \mbox{ad}_{\mathfrak{g}}}\end{equation}
The denominator is the subspace of parameters of the form $B^{ab} = f^{ab}_c A^c$. Such deformations were
shown to be BRST-trivial in \cite{Flores:2019dwr}.

To summarize, the resolution of the BRST trivial operator
\begin{equation}\Lambda\langle x\rangle \mbox{STr}(\lambda_L\lambda_R)\end{equation}
generated a cohomology class in:
\begin{equation}H^1\left(
         \mathfrak{g},
         \mbox{Hom}\left(
                         \mbox{ad}_{\mathfrak{g}},
                         {(\mbox{ad}_{\mathfrak{g}}\wedge\mbox{ad}_{\mathfrak{g}})_0\over \mathfrak{g}}
                         \right)
         \right)\end{equation}

This procedure generates the beta-deformation vertex out of simpler vertices
(the ghost number one vertex and the dilaton zero mode).
This example is a simpler version of Eq. (\ref{IntroSeqExt3}). We have a short exact sequence, {\it i.e.} just one intermediate
term, Eq. (\ref{ExampleShortExactSequence}) instead of Eq. (\ref{IntroSeqExt3}), and this is $\mbox{Ext}^1$:
\begin{equation}\mbox{Ext}_{\mathfrak{g}}^1\left(
                  \mbox{ad}_{\mathfrak{g}}
                  \,,\;
                  {(\mbox{ad}_{\mathfrak{g}}\wedge\mbox{ad}_{\mathfrak{g}})_0\over \mbox{ad}_{\mathfrak{g}}}
                  \right)\end{equation}

\section{Bicomplex}\label{Bicomplex}

From now on we will write, for brevity, $\mathfrak{g}$ instead of $\mbox{ad}_{\mathfrak{g}}$.

Descending from
$\Lambda\langle x\rangle\mbox{STr}(\lambda_L\lambda_R)\in\mbox{Coind}_H^G P^3$
to an element of  $\mbox{Ext}^1_{\mathfrak{g}}\left(\mathfrak{g}, {(\mathfrak{g}\wedge\mathfrak{g})_0\over \mathfrak{g}}\right)$
can be summarized in terms of the bicomplex which we will now describe, and its two spectral sequences.

For any representation $\cal R$, we can consider the double complex:
\begin{equation}C^{p,q} = C^p\left(\,
                   \mathfrak{g}
                   \,,\,
                   \mbox{Hom}\left(\,
                                   {\cal R}
                                   \,,\,
                                   \mbox{Coind}_H^GP^q
                                   \right)
                   \right)\end{equation}
with the differential
\begin{equation}d_{\rm tot} = d_{\rm Lie} + Q\end{equation}
We could first compute the BRST cohomology, and then the Lie superalgebra cohomology.
Or, we could first compute the Lie superalgebra cohomology and then BRST cohomology.
Correspondingly, there two spectral sequences, which we call $E$ and $\tilde{E}$.
The second page of the first spectral sequence is:
\begin{align}
E_2^{p,q}\;=\; & \mbox{Ext}^p_{\mathfrak{g}}
\left(\,
      {\cal R}
      \,,\,
      H^q(Q)
      \,\right)\label{SpecE} \\
d_2\;:\; & E_2^{p,q}\rightarrow E_2^{p+2,q-1}\end{align}
The second page of the second spectral sequence is:
\begin{align}
\widetilde{E}_2^{p,q}\;=\; & H_Q^p\left(
           \mbox{Ext}^q_{\mathfrak{g}}
           \left(\,
                 {\cal R}
                 \,,\,
                 \mbox{Coind}_H^G P^{\bullet}
                 \,\right)
           \right) \\
d_2\;:\; & \widetilde{E}_2^{p,q}\rightarrow \widetilde{E}_2^{p+2,q-1}\label{SpecEtil}\end{align}
Since both spectral sequences compute the cohomology of $d_{\rm tot}$,
\begin{equation}\bigoplus_{p+q=n}\widetilde{E}_{\infty}^{p,q} = \bigoplus_{p+q=n} E_{\infty}^{p,q}\end{equation}

Returning to
Section \ref{ToyExample},
our exact vertex
$\mbox{STr}(\lambda_3\lambda_1){\cal V}_1(u)$
fits into $\widetilde{E}_2^{3,0}$:
\begin{equation}\mbox{STr}(\lambda_3\lambda_1){\cal V}_1(u)
\;\in\;
\widetilde{E}_2^{3,0}
\;=\;
H^3_Q\left(
           \mbox{Hom}_{\mathfrak{g}}\left(\mathfrak{g},\mbox{Coind}_H^G P^{\bullet}
                                                   \right)
           \right)\end{equation}
while the extension cocycle $\tau$ belongs to $E_2^{1,2}$:
\begin{equation}\tau
\;\in\;
E^{1,2}_2
\;=\;
\mbox{Ext}^1_{\mathfrak{g}}\left(\mathfrak{g},H_Q^2(\mbox{Coind}_H^G P^{\bullet})\right)\end{equation}

\resizebox{0.7\textwidth}{!}{\input{"e2o.pgf"}}

\resizebox{0.7\textwidth}{!}{\input{"et2o.pgf"}}

\section{Finite dimensional representations in AdS/CFT}\label{FiniteDimensionalRepresentations}

The space of solutions of linearized SUGRA equations in $AdS_5\times S^5$ has infinitely many subspaces
forming finite-dimensional representations of $\mathfrak{g}$ \cite{Mikhailov:2011af}.
We conjecture that they are semisimple (the direct sum of irreducible representations).
The candidate irreducible representations can be described in terms of super-Young diagramms.
We will sketch their construction here, and then make it more precise in
Section \ref{SpaceOfDeformations}.

\subsection{Dilaton}\label{sec:Dilaton}

Let us start by considering the linearized equations for the dilaton in $AdS_5\times S^5$.
The solutions can be constructed by the embedding:
\begin{equation}AdS_5\times S^5 \;\subset\; \mathbb{R}^{2+4}\times \mathbb{R}^6\end{equation}
The dilaton satisfies the Laplace equation, and the solutions can be constructed as the restrictions of the
harmonic polynomials satisfying the following condition:
the degree in $\mathbb{R}^{2+4}$ should be equal to the degree in $\mathbb{R}^6$. Such polynomials are of the form:
\begin{equation}p(X,Y) = V^{i_1\ldots i_n}_{j_1\ldots j_n} X_{i_1}\cdots X_{i_n} Y^{j_1}\cdots Y^{j_n}\end{equation}
where $X_i$ are coordinates in $\mathbb{R}^6$ and $Y^j$ coordinates in  $\mathbb{R}^{2+4}$.

To supersymmetrize this, we must first translate into the spinor language, {\it i.e.} replace
$so(6)$ with $u(4)$ and $so(2,4)$ with $u(2,2)$.
In this language $X_i$ and $Y^j$ become antisymmetric tensors $X_{\alpha\beta}$ and $Y^{ab}$ with the
reality condition:
\begin{equation}X_{\alpha\beta} = {1\over 2}\epsilon_{\alpha\beta\gamma\delta}\overline{X}_{\gamma\delta}\end{equation}
and same for $Y$. Here we will discuss the finite-dimensional representations, where the reality conditions are
unimportant. Therefore we will replace both $u(2,2)$ and $u(4)$ with $gl(4)$, and relax the reality conditions
on $X$ and $Y$. In this language, the dilaton solution is:
\begin{equation}\label{DilatonSolution}p(X,Y) =
C^{a_1b_1\;a_2b_2\cdots a_nb_n}_{\alpha_1\beta_1\;\alpha_2\beta_2\cdots \alpha_n\beta_n}
X_{[a_1b_1]}\cdots X_{[a_nb_n]}\;Y^{[\alpha_1\beta_1]}\cdots Y^{[\alpha_n\beta_n]}\end{equation}
The Laplace operators in the ambient space are:
\begin{equation}\epsilon_{abcd}{\partial\over\partial X_{[ab]}}{\partial\over\partial X_{[cd]}}
\quad
\mbox{\tt and }
\quad
\epsilon^{\alpha\beta\gamma\delta}{\partial\over\partial Y^{[\alpha\beta]}}{\partial\over\partial Y^{[\gamma\delta]}}\end{equation}
Moreover, since we have to consider the restriction only, it is enough to consider the polynomials
modulo
\begin{equation}\epsilon^{a_1b_1a_2b_2}v^{a_3b_3\cdots a_nb_n}_{\alpha_1\beta_1\cdots\alpha_n\beta_n}\end{equation}
This implies that the dilaton solutions can be described with Eq. (\ref{DilatonSolution}) where
$C^{a_1b_1\;a_2b_2\cdots a_nb_n}_{\alpha_1\beta_1\;\alpha_2\beta_2\cdots \alpha_n\beta_n}$
is symmetric in all $a$-indices, all $b$-indices, all $\alpha$-indices and all $\beta$-indices.

\subsection{Supersymmetrization}\label{sec:Supersymmetrization}

What happens when we act on the dilaton solutions with the supersymmetries?
It was conjectured in \cite{Mikhailov:2011af} that the resulting finite-dimensional representation
of $psl(4|4)$ corresponds to some  super-Young diagram.
We will now describe the relevant Young symmetrizers.

We use the notations from \cite{MoensSchurFunctions}.

We can construct  finite-dimensional representations of $psl(4|4)$ from tensor product
of several fundamental and antifundamental representations. The main difference with the
representation theory of the classical Lie algebra $sl(N)$ is that the antifundamental representation
can not be constructed from the tensor products of fundamental representations.

What we do is to take a tensor product of several fundamental representations with several
antifundamental representations, and then apply some Young symmetrizer to the upper (fundamental) indices,
and another Young symmetrizer to lower (antifundamental) indices.

The finite-dimensional representations ${\cal H}_n$ are direct sums:
\begin{equation}{\cal H}_n = H_n \oplus H^{\star}_n\end{equation}
Let $F$ denote fundamental representation of $sl(4|4)$, and $F'$ antifundamental representation.
To obtain $H_n$, we take the tensor product of $2n$ fundamental and $2n$ antifundamental representations:
\begin{equation}v_1\otimes\cdots \otimes v_n\otimes w_1\otimes \cdots \otimes w_n\otimes
f^1\otimes\cdots\otimes f^n\otimes g^1\otimes \cdots\otimes g^n \in F^{\otimes 2n}\otimes (F')^{\otimes 2n}\end{equation}
Then, we do symmetrization or antisymmetrization over four $n$-tuples:
\begin{equation}v_{(1}\otimes\cdots\otimes v_{n)}\otimes w_{(1}\otimes \cdots\otimes w_{n)}\otimes
f^{[1}\otimes\cdots\otimes f^{n]}\otimes g^{[1}\otimes \cdots\otimes g^{n]}\end{equation}
Then we antisymmetrize each $v\leftrightarrow w$ pair and symmetrize each $f\leftrightarrow g$ pair.

\begin{figure}\centering\includegraphics[scale=0.45]{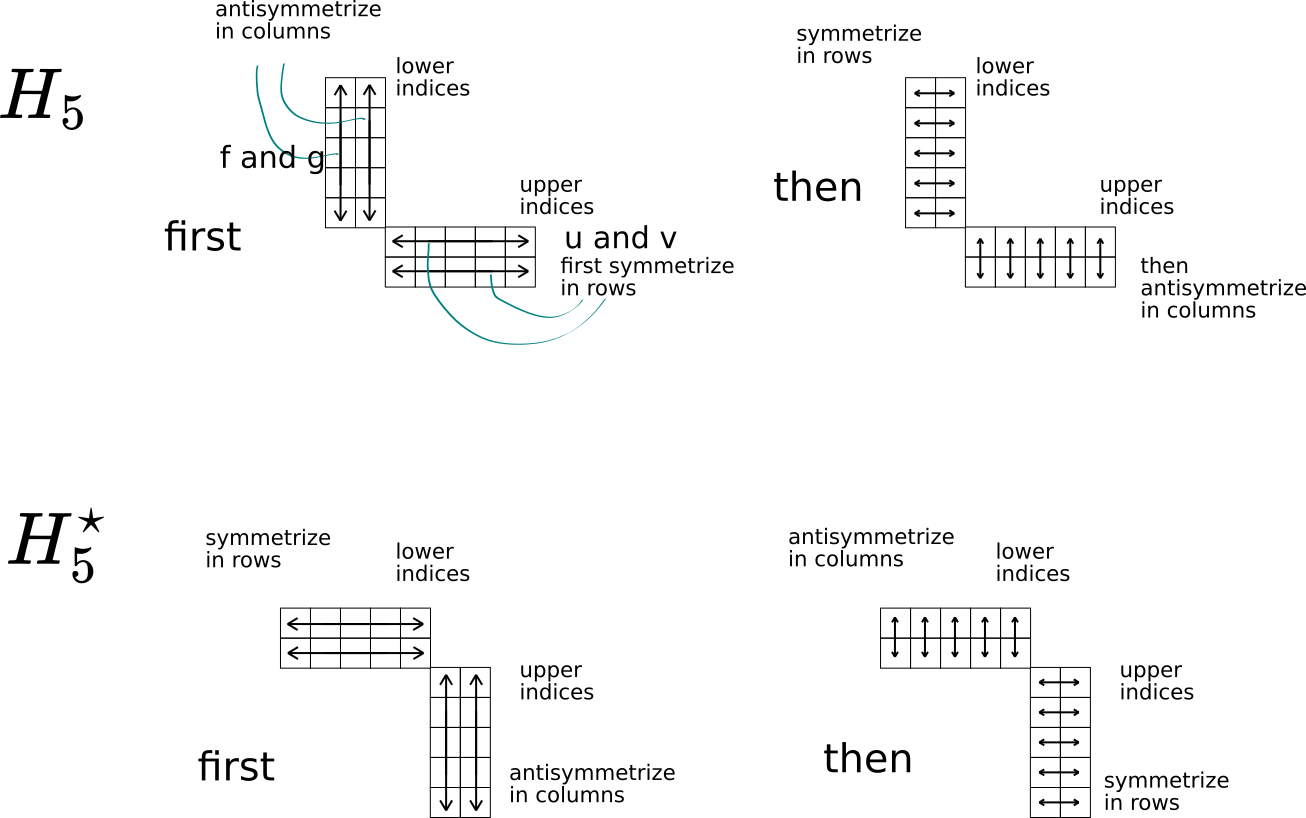}\caption{Young diagramms corresponding to $\int d^4x \;\mbox{tr}Z^7$}\end{figure}

For $H^{\star}$, both diagramms are transposed.
Notice that $H^{\star}$ can be characterized as the dual of $H$.

This is not the whole story, because we have to impose some constraints on traces of these tensors,
and factor out by a subspace. We will discuss this in
Section \ref{SpaceOfDeformations}.

\section{Other Young diagramms}\label{YoungDiagramms}

We introduce the notations for the linear spaces of symmetrized tensors, for example:

\begin{figure}\centering\includegraphics[scale=0.45]{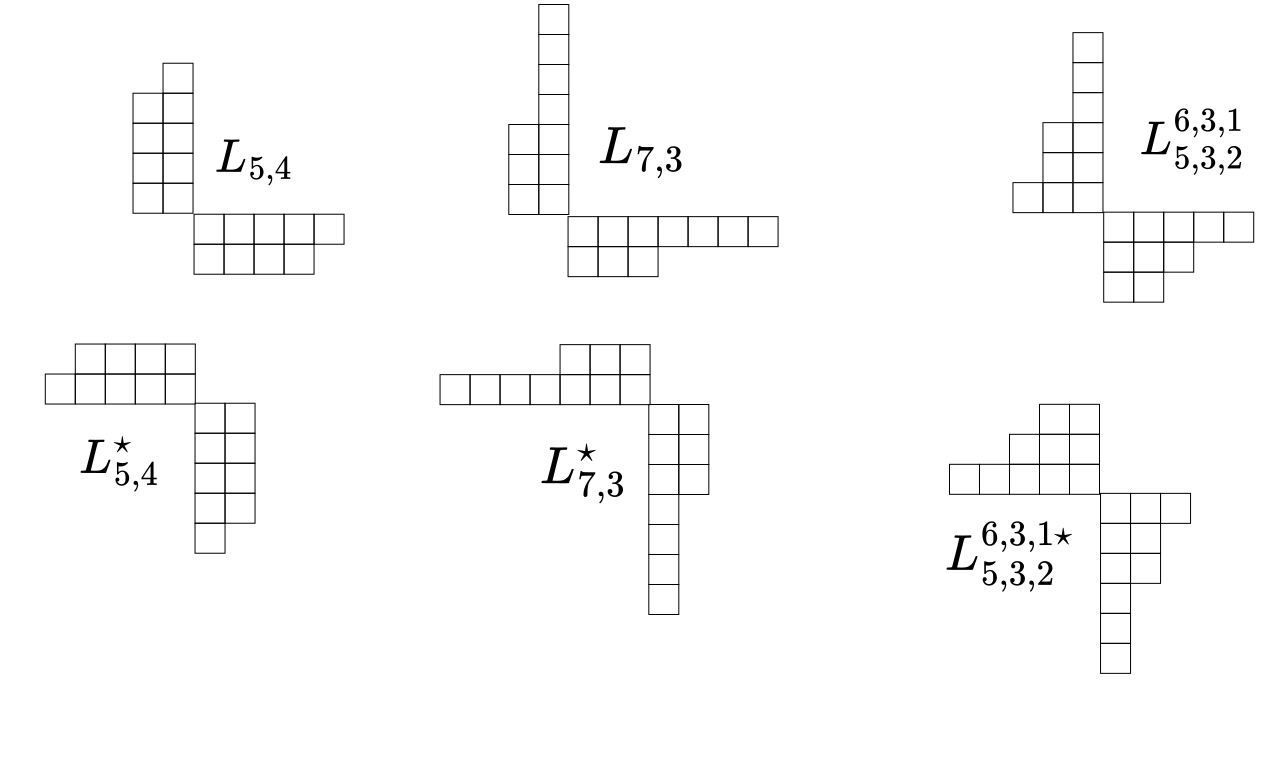}\caption{Some other Young diagramms}\end{figure}

These do not correspond to deformations of AdS, but will play a role.

\section{Trace maps and their duals}\label{TraceMaps}

There are very straightforward trace maps:
\begin{equation}s^{n+k-1,n}{}_{n+k,n}\;:\;L_{n+k,n}\rightarrow L_{n+k-1,n}\;,\;\; k\geq 1\end{equation}
and
\begin{equation}s^{n,n-1}_{n,n}\;:\;L_{n,n}\rightarrow L_{n,n-1}\end{equation}
This is just a trace over the corresponding indices, namely the rightmost vector of the longer row and
the topmost vector of the longer column.

\begin{figure}\centering\includegraphics[scale=0.5]{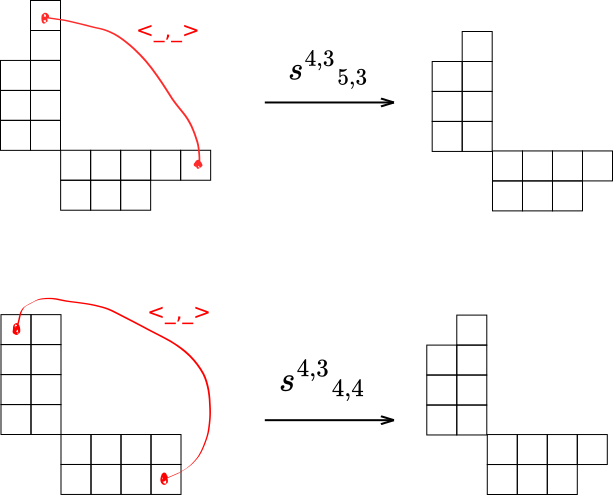}\caption{Super-trace map, lowering the rank of the tensor}\end{figure}

There is an exact sequence
\begin{align}
 & L_{n+1,n+1}\longrightarrow L_{n+1,n}\longrightarrow L_{n,n} \longrightarrow 0\end{align}
Similarly, there are maps
\begin{equation}s^{\star n+k-1,n}_{n+k,n}\;:\;L^{\star}_{n+k,n}\rightarrow L^{\star}_{n+k-1,n}\;,\;\; k\geq 1\end{equation}
and
\begin{equation}s^{\star n,n-1}_{n,n}\;:\;L^{\star}_{n,n}\rightarrow L^{\star}_{n,n-1}\end{equation}
which are defined in the same way.

Using the invariant scalar product, we define $\delta$ as the dual maps:
\begin{align}
\delta^{n+k,n}{}_{n+k-1,n}\;:\; & L_{n+k-1,n}\rightarrow L_{n+k,n} \\
\delta^{n+k,n}{}_{n+k-1,n}\;=\; & (s^{\star n+k-1,n}{}_{n+k,n})'\end{align}
The $\delta$-maps act as multiplication by the Kronecker delta symbol,
with subsequent symmetrizations-antisymmetrizations.

Notice that the $s$-maps are just contractions, no need to symmetrize-antisymmetrize afterwards.
After the contraction, the resulting tensor will already have the right symmetry type.
There are other trace maps, for example $L_{n+k,n}\rightarrow L_{n+k,n-1}$ which do require
the subsequent application of the Young projector. But those which shorten either the longest row/column,
or one of two equal length, do not require  additional projectors.

{\bf Lemma \refstepcounter{Theorems}\label{S2}\noindent{\bf \arabic{Theorems}}}:
\begin{equation}(s^2\;:\;L_{n,n} \rightarrow L_{n-1,n-1}) = 0\end{equation}
This $s^2$ is the composition of $s\;:\;L_{n,n}\rightarrow L_{n,n-1}$ and $s\;:\;L_{n,n-1}\rightarrow L_{n-1,n-1}$.

{\bf Lemma \refstepcounter{Theorems}\label{D2}\noindent{\bf \arabic{Theorems}}}:
\begin{equation}(\delta^2\;:\;L_{n,n} \rightarrow L_{n+1,n+1}) = 0\end{equation}
This follows from Lemma \ref{S2}, since $\delta$ is dual to $s$.

{\bf Lemma \refstepcounter{Theorems}\label{DSSD0}\noindent{\bf \arabic{Theorems}}}
Consider the following maps:
\begin{equation}L_{n+1,n}
{\overset{s}{\rightarrow}\atop\underset{\delta}{\leftarrow}}
L_{n,n}
{\overset{s}{\rightarrow}\atop\underset{\delta}{\leftarrow}}
L_{n,n-1}\end{equation}
With these notations:
\begin{equation}\label{SDeltaIsDeltaS}s^{n,n}{}_{n+1,n}\delta^{n+1,n}{}_{n,n} =    \delta^{n,n}{}_{n-1,n} s^{n-1,n}{}_{n,n}\end{equation}
up to a numerical factor.

\begin{figure}\centering\includegraphics[scale=0.5]{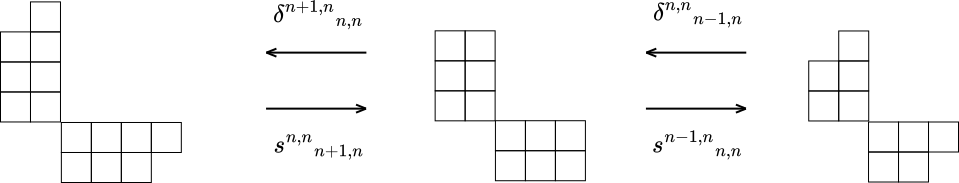}\caption{Eq. (\ref{SDeltaIsDeltaS})}\end{figure}

\section{Space of deformations}\label{SpaceOfDeformations}

Finite dimensional deformations have been discussed in \cite{Mikhailov:2011af}.
We will now formulate the conjecture more precisely.
The space of deformations, which we will call $H_n$, is obtained from $L_{n,n}$
by a restriction to a subspace and taking a factorspace, as we now describe.

\subsection{Definition of $H_n$}\label{DefH}

Let us define the subspace:
\begin{align}
L_{n,n}^0 \;\subset\; & L_{n,n} \\
L_{n,n}^0 \;=\; & \mbox{ker}\;s^{n,n-1}{}_{n,n} \\
(\delta L_{n,n-1})^0\;\subset\; & L_{n,n}^0 \\
(\delta L_{n,n-1})^0\;=\; & L_{n,n}^0\cap \delta^{n,n}{}_{n,n-1}L_{n,n-1} \\
L_{n,n-1}^{00}\;\subset\; & L_{n,n-1} \\
L_{n,n-1}^{00}\;=\; & \mbox{ker} \left(s^{n-1,n-2}{}_{n-1,n-1}s^{n-1,n-1}{}_{n,n-1}\right)\end{align}
We define:
\begin{equation}H_n = {L_{n,n}^0\over (\delta L_{n,n-1})^0}\end{equation}
Let us also consider the subspace:
\begin{align}
 & D_n \subset L_{n,n-1}\label{DefDn} \\
 & D_n = \delta L_{n-1,n-1} + \delta L_{n,n-2} + \delta L_{n,n-2}^{n-1,n-1} + \delta L_{n-1,n-1}^{n,n-2}\end{align}
This is the space of all tensors containing the Kronecker delta-symbol;
it is the direct sum of four invariant subspaces.
Notice that:
\begin{equation}\label{sDn}sD_n = \delta L_{n-1,n-2}\end{equation}

The space $H^{\star}_n$ is defined similarly.

\subsection{Scalar product}\label{sec:ScalarProduct}

There is an invariant coupling (scalar product):
\begin{equation}H^{\star}_n\otimes H_n \longrightarrow \mathbb{C}\end{equation}
We have to verify that the scalar pairing $\langle v^{\star},v\rangle$ respects
the equivalence relation $v\simeq v + \delta u$. Indeed:
\begin{equation}\langle v^{\star}, \delta u \rangle = \langle s^{\star}v^{\star}, u\rangle = 0\end{equation}
This pairing is nondegenerate, since $H_n$ is an irreducible representation.
This implies that $H^{\star}_n$ is the dual of $H_n$:
\begin{equation}H_n^{\star} = H'_n\end{equation}

\subsection{Exact sequences involving $H_n$}\label{sec:ExactSequencesWithH}

\subsubsection{First exact sequence involving  $H_n$}\label{sec:FirstExactSequence}

\begin{equation}\label{FirstSequence}0
\longrightarrow
H_n
\longrightarrow
{L_{n,n}\over \delta L_{n,n-1}}
\overset{s^{n,n-1}{}_{n,n}}{\longrightarrow}
{L_{n,n-1}^{00}\over D_n^{00}}
\overset{s^{n-1,n-1}{}_{n,n-1}}{\longrightarrow}
{L_{n-1,n-1}^0\over sD_n^{00}}
\longrightarrow
0\end{equation}
where
\begin{equation}D_n^{00} = D_n\cap L_{n,n-1}^{00}\end{equation}
Eq. (\ref{sDn}) implies that $sD_n^{00}=(\delta L_{n,n-1})^0$. Therefore ${L_{n-1,n-1}^0\over sD_n^{00}} = H_{n-1}$ and
Eq. (\ref{FirstSequence}) is defines a cohomology class:
\begin{equation}\label{PsiDown}\Psi_{n-1}^n \in \mbox{Ext}^2(H_{n-1},H_n)\end{equation}
Similar considerations with $L^{\star}$ instead of $L$ define:
\begin{equation}\Psi^{\star}{}_{n-1}^n \in \mbox{Ext}^2(H^{\star}_{n-1},H^{\star}_n)\end{equation}

\subsubsection{Second exact sequence involving $H_n$}\label{sec:SecondExactSequence}

Using the scalar product, we can construct the dual sequences:

\begin{align}
 & 0
\longrightarrow
H^{\star}_{n-1}
\longrightarrow
\left({L_{n,n-1}^{00}\over D_n^{00}}\right)'
\longrightarrow
\left({L_{n,n}\over \delta L_{n,n-1}}\right)'
\longrightarrow
H^{\star}_n
\longrightarrow
0 \\
 & 0
\longrightarrow
H_{n-1}
\longrightarrow
\left({L_{n,n-1}^{\star 00}\over D_n^{\star 00}}\right)'
\longrightarrow
\left({L^{\star}_{n,n}\over \delta L^{\star}_{n,n-1}}\right)'
\longrightarrow
H_n
\longrightarrow
0\end{align}
These sequences define:
\begin{align}
 & \Psi^{\star}{}_n^{n-1} \in \mbox{Ext}^2(H^{\star}_n, H^{\star}_{n-1}) \\
 & \Psi_n^{n-1} \in \mbox{Ext}^2(H_n, H_{n-1})\label{PsiUp}\end{align}

\section{Non-covariance of ghost number three vertex operators}\label{GhostNumberThree}

The ghost number three cohomology is nontrivial \cite{Mikhailov:2014qka}.
One can construct ghost number three vertices by multiplying ghost number
two vertices by the ghost number one vertices corresponding to the global symmetries.
The ghost number three cohomology transforms in the same representation as the ghost number two cohomology.
However, the symmetry transformations of the corresponding {\it vertices} are more complicated.
Unlike the case of ghost number two, for the ghost number three there {\it is}, actually,
an obstacle for the vertices to transform covariantly.

Let ${\cal H}\subset H_Q^2$ be an invariant subspace. The action of $\mathfrak{g}$ defines a map
\begin{align}
a\;:\; & \mathfrak{g}\otimes {\cal H} \longrightarrow {\cal H} \\
a(x\otimes v)\;=\; & a.v\end{align}
As we explained in Section \ref{SpaceOfDeformations},
$H$ has a scalar product.
Therefore, we can consider the dual map:
\begin{equation}a^{\dagger}\;:\; {\cal H}\longrightarrow \mathfrak{g}\otimes {\cal H}\end{equation}
The second Casimir vanishes on linearized SUGRA solutions, therefore:
\begin{equation}aa^{\dagger} = 0\end{equation}
Therefore we can consider the  exact sequence:
\begin{align}
 & 0\longrightarrow
{
 \mbox{ker}\;(\mathfrak{g}\otimes {\cal H}\overset{a}{\rightarrow} {\cal H})
 \over
 \mbox{im}\;({\cal H}\overset{a^{\dagger}}{\rightarrow} \mathfrak{g}\otimes {\cal H})
 }
\longrightarrow
{\mathfrak{g}\otimes {\cal H}\over \mbox{im}\;({\cal H}\overset{a^{\dagger}}{\rightarrow} \mathfrak{g}\otimes {\cal H})}
\longrightarrow {\cal H}\longrightarrow 0\end{align}
and the corresponding cohomology class
\begin{equation}\omega\;\in\;
\mbox{Ext}^1
\left(
      \;{\cal H}
      \;,\;
      {
      \mbox{ker}\;(\mathfrak{g}\otimes {\cal H}\overset{a}{\rightarrow} {\cal H})
      \over
      \mbox{im}\;({\cal H}\overset{a^{\dagger}}{\rightarrow} \mathfrak{g}\otimes {\cal H})
      }
      \right)\end{equation}
The space
\begin{equation}\mbox{ker}\;(\mathfrak{g}\otimes {\cal H}\overset{a}{\rightarrow} {\cal H})
\over
\mbox{im}\;({\cal H}\overset{a^{\dagger}}{\rightarrow} \mathfrak{g}\otimes {\cal H})\end{equation}
has an invariant scalar product. Therefore, we can consider the dual exact sequence:
\begin{equation}0\longrightarrow {\cal H}\overset{a^{\dagger}}{\longrightarrow}
\mbox{ker}\;(\mathfrak{g}\otimes {\cal H} \overset{a}{\longrightarrow} {\cal H})
\longrightarrow
{
 \mbox{ker}\;(\mathfrak{g}\otimes {\cal H}\overset{a}{\rightarrow} {\cal H})
 \over
 \mbox{im}\;({\cal H}\overset{a^{\dagger}}{\rightarrow} \mathfrak{g}\otimes {\cal H})
 }
\longrightarrow 0\end{equation}
and the corresponding cohomology class
\begin{equation}\omega'\;\in\;
\mbox{Ext}^1
\left(
      {
      \mbox{ker}\;(\mathfrak{g}\otimes {\cal H}\overset{a}{\rightarrow} {\cal H})
      \over
      \mbox{im}\;({\cal H}\overset{a^{\dagger}}{\rightarrow} \mathfrak{g}\otimes {\cal H})
      }
      \;,\;
      \;{\cal H}
      \right)\end{equation}
The extension $\omega$ describes an ambiguity of representing $v\in {\cal H}$ as some $x.w$,
{\it i.e.} finding a ghost number three vertex operator corresponding to $v$.
On the other hand, we conjecture that $\omega'$  corresponds to the resolution
of BRST exact ghost number three vertices of the form $\Lambda\langle x\rangle V\langle w\rangle$ where
$x.w=0$. A prerequisite of this conjecture is that the vertices in $\mbox{im}\;a^{\dagger}$ are
resolved covariantly;
we do not know if this is true.

If this conjecture is true, then the Yoneda product:
\begin{equation}\omega'\omega\;\in\; \mbox{Ext}^2\left(H,H\right)\end{equation}
is the obstacle to the covariance of the ghost number three vertices.

In fact, there are other classes in $\mbox{Ext}^2\left(H,H\right)$, namely
$\Psi^n_{n+1}\;\in\;\mbox{Ext}^2\left(H_{n+1},H_n\right)$
and
$\Psi^{n+1}_n\;\in\;\mbox{Ext}^2\left(H_n,H_{n+1}\right)$
constructed in
Section \ref{SpaceOfDeformations}.
They could potentially be obstacles as well.

\section{Resolving wedge products}\label{ResolvingWedgeProducts}

We denote $F$  fundamental and $F'$ antifundamental representation of $sl(4|4)$.

Consider the subspace $S^n F\subset F^{\otimes n}$ of symmetric tensors, and
subspace $\Lambda^nF'\subset (F')^{\otimes n}$ of antisymmetric tensors.

Consider the following two linear maps:
\begin{align}
\delta\;:\; & S^n F\otimes \Lambda^n F' \rightarrow S^{n+1} F\otimes \Lambda^{n+1} F' \\
s\;:\; & S^{n+1} F\otimes \Lambda^{n+1} F' \rightarrow S^n F\otimes \Lambda^n F'\end{align}
The map $\delta$ is the tensor multiplication by ${\bf 1}\subset F\otimes F'$, with subsequent
symmetrization/antisymmetrization.
The map $s$ is taking the super-trace, {\it i.e.} the pairing $F'\otimes F \rightarrow \mathbb{C}$.
Their anticommutator $\{s,\delta\}$ is zero:
\begin{equation}\{s,\delta\} = 0\end{equation}
We denote:
\begin{equation}K_n = {(S^n F\otimes \Lambda^n F')^0\over\mbox{im}\;\delta}\end{equation}
where the index $(\ldots)^0$ means restriction to $\mbox{ker}\;s$.

Consider the space of covariant cohomologies transforming in $K_n$, of ghost number $n$.
In terms of the spectral sequence of Eq. (\ref{SpecEtil}):
\begin{equation}\widetilde{E}_2^{n,0} = H^n_Q\left(\mbox{Hom}_{\mathfrak{g}} (K_n, \mbox{Coind}_H^G P^{\bullet})\right) = \mathbb{C}\end{equation}
It is generated by
\begin{equation}\label{ExactWedgeProduct}{\cal A}\mapsto \left\langle {\cal A}\;,\; g^{-1}(\lambda_L-\lambda_R)g \otimes\cdots\otimes g^{-1}(\lambda_L-\lambda_R)g\right\rangle\end{equation}
Notice that this the cohomology class of the covariant complex, not just covariant vertex.
Covariant vertex by itself would not involve the factorizaion of the space of $\cal A$ by $\delta(\ldots)$.
We need to impose the tracelessness condition, because $g^{-1}(\lambda_L-\lambda_R)g$ is only well-defined
up to a unit matrix.

This expression for $n>3$ is $Q$-exact in the space non-covariant $\mbox{Coind}_H^GP^{\bullet}$.
The resolution gives nontrivial squares  in the spectral sequence of Eq. (\ref{SpecE}):
\begin{align}
E_2^{n-3,3}=\mbox{Ext}^{n-3}\left(K_n,H_Q^3\right) & \mbox{conjectured zero} \\
E_2^{n-2,2}=\mbox{Ext}^{n-2}\left(K_n,H_Q^2\right) &  \\
E_2^{n-1,1}=\mbox{Ext}^{n-1}\left(K_n,H_Q^1\right) &  \\
E_2^{n,0} = \mbox{Ext}^{n}\left(K_n,\mathbb{C}\right) & \end{align}
The natural conjecture is that the only nonzero one is $\mbox{Ext}^{n-2}\left(K_n,H_Q^2\right)$, given
by the following exact sequence:
\begin{align}
0\longrightarrow & {(S^2 F\otimes \Lambda^2 F')^0
\over \mbox{im}\;\delta}
\stackrel{\delta}{\longrightarrow}
(S^3 F\otimes \Lambda^3 F')^0
\stackrel{\delta}{\longrightarrow}
\ldots \\
\ldots\stackrel{\delta}{\longrightarrow} & (S^n F\otimes \Lambda^n F')^0
\stackrel{p}{\longrightarrow}
{(S^n F\otimes \Lambda^n F')^0\over \mbox{im}\;\delta}
\longrightarrow
0\end{align}
This defines an element
\begin{equation}c \in
\mbox{Ext}^{n-2}\left(K_n,K_2\right)\end{equation}
Notice that $K_2$ is a subspace of $H^2_Q$, namely the space of beta-deformations (including the nonphysical ones).
The  natural conjecture is that $c$ corresponds to the resolution of the BRST exact expressions of the form Eq. (\ref{ExactWedgeProduct}).

\section{Resolving $\mathfrak{g}$-invariant vertices}\label{ResolvingScalars}

Consider the resolution of $\mathfrak g$-invariant operators, such as:
\begin{equation}\mbox{STr}(\lambda_3\lambda_1)^n\end{equation}
Using the notations of
Section \ref{Bicomplex}
the space of such operators is $\widetilde{E}_2^{2n,0}$ for ${\cal R}=\mathbb C$.

We conjecture that the resolution of an invariant polynomial of the degree $2n$ results in a cohomology
class of the form of the Yoneda product of $n-1$ classes $\Psi^k_{k\pm 2}$ defined in Eqs. (\ref{PsiUp}) and (\ref{PsiDown}):
\begin{equation}\label{Ext0M}\Psi^m_{k_{n-2}}\Psi^{k_{n-2}}_{k_{n-3}}\cdots\Psi^{k_1}_0
\in
\mbox{Ext}^{2n-2}\left(\mathbb{C}, H_m\right)\end{equation}
Consider a ghost number two vertex operator representing a state in $H_p\subset H^2_Q$.
Let us multiply it by an invariant polynomial of $\lambda_L$ and $\lambda_R$ of the degree $2n$.
This will be a BRST exact vertex. We conjecture that its resolution corresponds to the cohomology classes of the
form
\begin{equation}\label{ExtPQ}\Psi^q_{k_{n-1}}\Psi^{k_{n-1}}_{k_{n-2}}\cdots\Psi^{k_1}_p
\in
\mbox{Ext}^{2n}\left(H_p, H_q\right)\end{equation}

\section{Conclusion and open questions}\label{OpenQuestions}

Vertex operators with high ghost number are BRST trivial, and their resolution (finding $W$ such that $V = QW$)
should play an important role in the computations. In this paper we have shown that the resolution naturally defines
some cohomology classes, which classify exact sequences. After the thorough investigation of the simplest example
in Section \ref{ToyExample}, we proceed to develop the general formalism.
We concentrate in the finite dimensional representations previously studied in \cite{Mikhailov:2011af}, and
show that they fit into some 4-term exact sequences (corresponding to $\mbox{Ext}^2$),
Section \ref{SpaceOfDeformations}. This allows us to identify
the cohomology classes which should correspond to the resolutions of some families of high ghost number operators, in
Section \ref{ResolvingWedgeProducts}
and
Section \ref{ResolvingScalars}.
Namely, we consider operators like $\mbox{STr}(\lambda_L\lambda_R)^N$ (and other $\mathfrak{g}$-invariant operators)
and $(\lambda_L-\lambda_R)^{\wedge N}$ (the higher ghost number BRST-trivial generalization of the beta-deformation vertex).
In both cases, we discover the candidate Ext classes and write the corresponding exact sequences.

We will now point out some open questions.

\subsection{Proving the conjecture about the structure of finite-dimensional representations}\label{sec:ProveFiniteDimensional}

It has been conjectured in \cite{Mikhailov:2011af} that the finite-dimensional representations
of linearized SUGRA solutions are irreducible. (More precisely, they  split as direct sums of irreps.)
If they are not irreducible, this will make the story even more interesting.
This is, in principle, a straightforward question. On the field theory side, take a deformation, for example:
\begin{equation}\int d^4x \mbox{Tr} Z^{10}\end{equation}
and act on it by all  superconformal transformation. Will the resulting representation be irreducible?
The splitting property  is probably related to the existence of the scalar product.
The spaces $H_n \oplus H_n^{\star}$ have an invariant scalar product
Section \ref{sec:ScalarProduct}.
On the field theory side, this must be defined as the correlation function of two deformations.
However, since the deformations overlap, there are potential divergencies from the collision of the
integration points. For example, consider the correlation function:
\begin{equation}\left\langle
\int d^4x \mbox{Tr} Z^{10}(x)
\int d^4y \mbox{Tr} \overline{Z}^{10}(y)
\right\rangle\end{equation}
there is a divergence when $x$ and $y$ collide. This requires a regularization.
Therefore, the question of  existence of an invariant scalar product requires further examination.

Also, it is necessary to actually prove  that the exact sequences constructed in
Section \ref{SpaceOfDeformations}
are nontrivial (do not split). We do not have a proof that Eqs. (\ref{Ext0M}) and (\ref{ExtPQ}) are nonzero in cohomology.

\subsection{Include infinite-dimensional representations}\label{sec:IncludeInfiniteDim}

Why would the resolution of  $(\mbox{STr}(\lambda_L\lambda_R))^N$ only give rise to finite-dimensional representations?
This seems to be a reasonable conjecture, but we  do not have any argument towards its validity.
(Other than this being the simplest possibility.)
In any case, it is an open problem to extend the formalism of this paper to include infinite-dimensional representations.

\subsection{Actual computations of the cocycles}\label{sec:DevelopComputations}

The computations are, in principle, completely straightforward. Given a cocycle with high enough ghost number
(such as $(\mbox{STr}(\lambda_L\lambda_R))^N$ with $N>1$), we know that it is BRST-exact. We can use an explicit
formula for the BRST operator, in local coordinates, and find the explicit expression for $Q^{-1}(\mbox{STr}(\lambda_L\lambda_R))^N$
order by order in the coordinate expansion. This calls for a symbolic computation. Unfortunately, straightforward symbolic
computations of the BRST complex in $AdS_5\times S^5$ seems to be unfeasible, mostly due to the very large
number (32) of odd coordinates. Perhaps a collection of skillful computational tricks, along the lines of
\cite{Fleury:2021ieo}, would work.

\section{Acknowledgments}\label{Acknowledments}

This work was supported in part by  FAPESP  grant 2019/21281-4,
and in part by FAPESP grant 2021/14335-0,
and in part by CNPq grant ``Produtividade em Pesquisa'' 307191/2022-2.

\def\cprime{$'$} \def\cprime{$'$}
\providecommand{\href}[2]{#2}\begingroup\raggedright\endgroup
\end{document}